# Assessment of Microbial Habitability Across Solar System Targets

Dimitra Atri[1,2], Todd Godderidge[3], Dee Cirium[4], Dimple Patel[5], Gunasekar Ramakrishnan[6]

[1] New York University Abu Dhabi, Center for Space Science, Abu Dhabi, United Arab Emirates (atri@nyu.edu)
[2] Blue Marble Space Institute of Science, Seattle, WA, USA
[3] Warwick University, Physics Department, Coventry, United Kingdom
[4] Imperial College London, Natural Sciences, London, United Kingdom
[5] National Institute of Technology Calicut, School of Biotechnology, Calicut, Kerala, India
[6] Muthurangam Government Arts College, Department of Physics, Vellore, Tamil Nadu, India

## Abstract

With a fleet of exploratory space missions on the horizon, the study of target specific biospheres is crucial for accurately determining the probability of the existence of microbial life on various planetary bodies and prioritising targets accordingly. Although previous studies have compared the potential habitability of objects in our solar system by bulk characteristics, it is less common that precise qualitative methods are developed for ranking candidates hospitable to microbial life on a local environment basis. In this review we create a planetary environmental database and use it to motivate a list of primary habitability candidates and essential criteria for microbial survival. We then propose a new method, the Microbial Habitability Index (MHI) which uses a metric of microbial survival factor values in target environments compared with appropriate Earth analogues to assess their potential for life. We arrive at a selection of eight primary candidates and from this set conclude that Europa, Mars, and Enceladus have the highest potential for facilitating microbial survival.

## 1. Introduction

The search and postulation of life beyond the confines of our planet have been ongoing for centuries. In the last few decades, there have been numerous space missions to collect information about the environments of different objects in the solar system and researchers have been evaluating the possibilities of the existence of microbial life in particular. Many studies have compiled comprehensive reviews about specific environmental conditions of various icy objects in our solar system and likewise have also explored the possible existence of microbial life in those environments. However, there have been few comprehensive environment-oriented studies (Merino et al. 2019) incorporating Earth analogue environments to compare with planetary objects and a systematic method to rank them as astrobiology candidates on a local environmental basis. Hence, mediated by the available data on extremophiles and environmental conditions on other worlds, we have developed a method; a metric for habitability and an indexing system.

Extremophiles present in different environments on Earth give us valuable insights on the possibility of finding life on other celestial bodies. Life as we know it requires the availability of three major components: a source of energy, a liquid solvent and chemical building blocks (Merino et al. 2019). Many of the known extremophiles are also polyextremophiles, surviving at the extreme limits of multiple extrinsic factors, such as temperature and pressure, simultaneously. We purport that these polyextremophiles define the known limits of life, and the aforementioned factors are key to understanding why. The major criteria for habitability of these extremophiles more generally are: the presence, persistence and chemical activity of liquid water, thermodynamic disequilibrium providing suitable energy sources, environmental factors that bear on the stability of covalent and hydrogen bonds in biomolecules (temperature, pH, salinity, irradiation), presence of bio-essential elements (C, H, O, N, P, S) and endogenic factors, including cosmic dangers. Through research on Earth-bound extremophiles, it has become evident that the limit of life varies when organisms face co-occurring multiple extremes. Thus, we aim to use this data and extrapolate for similar environments on other planets.

In this study, we conduct a literature review to motivate a list of primary habitability candidates and essential criteria for microbial survival (habitability/survival factors). Furthermore, we compile a dataset of specific environments selected to best represent the diversity of conditions within seven planetary bodies, which are among the most popular and explored of our habitability candidates and analyse them against Earth analogues through the lens of microbial habitability. Our Earth analogues have been chosen as representatives of the most extreme habitable biospheres. These biospheres, of which we find near parallels on our habitability targets are: surface – deserts and surface ices, subsurface – deep continental crust and base of ice crust, hydrosphere – oceans, deep sea floor and hydrothermal vents. However, not all environments are present on each target body, and the similarity of environments against the habitability criteria have been assessed. We then designed a new method to calculate the Microbial Habitability Index (MHI), which uses a metric of microbial survival factor values in target environments compared with appropriate Earth analogues to assess their potential for life. Combining this analysis with considerations of other endogenic factors which cannot be numerically input into such a metric, we arrive at a final ordering of main astrobiological targets, ranked in order of our proposed habitability prospects.

# 2. Background

# 2.1.1. Definitions of Microbial Habitability

Habitability is a conservative term describing whether a particular extra-terrestrial environment can sustain life based on the known limits of life on Earth. While the presence of life in a designated environment is not a prerequisite for habitability, considerations are given to whether an environment can generate or sustain life, timescales of sustenance and stability, the spatial distributions, and the metabolic state of lifeforms. Depending on the particular combination of these criteria, requisite conditions for life vary, and this has subsequently given rise to a host of specific definitions of habitability (Cockell et al. 2016). We adopt the following definition in our study: A fixed set of extrinsic macroscopic conditions in a given environment, across recent (observable) timescales, which would support the metabolic activity of microbial life while such conditions were maintained. To summarise, the four main limiting considerations we made were: Generation versus sustenance (wariness of the habitability bottlenecks, e.g. abiogenesis), spatial extent (planet-wide vs local restricted environments), timescales (continuous planetary habitability and instantaneous habitability, in this work we include the latter) and the dependence of preservation on metabolic state. We lay out more details on these specific Habitability Considerations below.

### Consideration 1 - Generation vs Sustenance:

In our understanding of the origin and evolution of life, it is widely recognised that a broad set of prebiotic conditions must enter a 'habitability bottleneck' in order for life to evolve, before once again broadening to the conditions occupied by known life (Cockell et al. 2016). Subsequently considerations of Abiogenesis restrict searches for life, and optimistic parameterisation will consider only the sustenance of life, e.g. through Forward Contamination.

### Consideration 2 - Spatial Extent:

Studies on Planet–Wide habitability consider a number of exogenic and endogenic factors which might serve to aid or inhibit life (e.g. size, solar distance, tidal locking). However, they do not fully represent the diversity of environments on a planet, which might support life despite inhospitable planet-wide conditions (e.g. crater brines, subsurface oceans, sulphur rich atmospheres). These Restricted Environments would better reflect endmember conditions and improve habitability estimates (Preston and Dartnell, 2014; Cockell et al. 2016; Merino et al. 2019).

### Consideration 3 - Timescales:

The generation, evolution and sustenance of life operate on different timescales with the endmembers instantaneous and continuous. Continuous Planetary Habitability refers to the sustenance of life on geological timescales and incorporates the coevolutionary relationship between life and planetary chemistry. This relationship is observed on Earth, where biological processes have enhanced surface disequilibria and allowed the proliferation of life on planet–wide scales (Watson, 1999; Cockell et al. 2016). Instantaneous Habitability relates to the conditions of an environment at a given time without consideration of cumulative history. While generation is represented by an instantaneous time in the past, it is rarely decoupled from evolution, which is a long-term time–transgressive process; hence abiogenesis is linked to continuous habitability. Sustenance, conversely, can be described on various timescales spanning the continuous–instantaneous range.

### Consideration 4 - Metabolic State:

The limits of life are defined by microbial extremophiles which operate on three levels: 1) Preservation: where microbes are in a stasis state with no metabolic activity, but retain their potential for revival to metabolically active states (e.g. via cryopreservation (Pikuta et al. 2009)); 2) Tolerance: where microbes are in a stressed environment which supports metabolic activity but not reproduction, i.e. a sub–optimal survival state; and 3) Propagation: where microbes are in an optimal environment where conditions are hospitable enough to support metabolic activity and reproduction (Cockell et al. 2016; Takai, 2019).

# 2.1.2 Habitability Criteria

There are several delineated criteria that must be satisfied for planetary habitability. Foremost of these are the presence (and activity) of liquid water, and any source of thermodynamic disequilibrium necessarily providing the energy for biochemical reactions to occur (Hays et al. 2015). Additionally, the importance of extrinsic environmental conditions has also been highlighted by e.g. (Cockell et al. 2016), as well as the abundance of bio–essential elements at the spatial scales of an organism. To this list we also add the categories of long term, time sensitive phenomenon, as well as endogenic and exogenic factors. For a suitably comprehensive list of habitability criteria which fall into these categories, we turn to previous work and leaders in this area, including the assessment of key factors conducted in the NASA Astrobiology Strategy. More information on the background and choice of these categories is included below.

Factor 1 - Liquid Water:
Water is the solvent in which biochemical reactions occur; its presence, persistence and activity are necessary to life. Liquid water is a function of temperature, pressure and chemical impurities; however not all of the liquid phase space is habitable to life.

Factor 2 - Disequilibrium Energy Sources:
Thermodynamic disequilibrium results in the presence of oxidants and reductants, which can be used to produce energy in biochemical reactions (redox chemistry being universal to life). On planetary scales, disequilibria are associated with geochemical turnover (e.g. resurfacing) and physical processes (Factors 4 and 5). Disequilibria can be highly varied between environments and are often concentrated at material interfaces (e.g. water–rock interactions).

Factor 3 - Bio-essential Elements:
The presence of bio–essential elements (major elements CHONPS and various metals on the periodic table e.g. K, Fe, Mg) are necessary to build biological macromolecules (lipids, sugars, proteins, genetic material). Linked to redox species in Factor 2, as well as planetary composition and evolution (Factors 4–6).

Factor 4 - Endogenic Factors:
Endogenic factors include the presence of atmospheres, continents and oceans; geochemical cycles; energy flux, tectonics (plate, stagnant lid) and drivers (solar, tidal, radiogenic); size; mass; composition; rotational period; obliquity; radiation environment; tectonic activity; stability and evolution.

<u>Factor 5 - Exogenic Factors:</u>
Exogenic factors include the stellar neighbourhood, host star, characteristics of the planetary system, volatility of the protoplanetary disk and bombardment history.

<u>Factor 6 - Timescales:</u>
Over the lifetime of a target, many bulk properties can change, as well as the conditions in specific environments, to such an extent as to transform into completely new environments. We consider timescales over which the above factors can be assumed to be consistent, and the environments we select are to be stable over such timescales. In this work we assess current habitability of targets as opposed to historic potential for habitability.

From this evaluation we finalised a list of twenty essential criteria for survivability which fell into the six categories detailed above: Activity of Water, Salinity, Ph and Eh, Transport (Inflow/outflow), Oxidants, Toxin Abundances, Reductants, Bio-Essential Elements, Solar Flux, Temperature, Desiccation (Relative Humidity), Pressure, UVC irradiation, Ionising radiation, Carbon Dioxide count, Temperature fluctuations (K/yr), Tectonic Activity, Tidal Forces and Chemical Gradients. In our study we focus on environmental conditions, which have been numerically constrained in various regional–scale environments on solar system bodies from space missions and models (Merino et al. 2019 and therein referenced). Our analyses will be correlated to qualitative studies conducted on energy sources, based on environmental analogues on Earth (Preston and Dartnell, 2014; Cockell et al. 2016). It is worth noting then, that this list was subject to filtering based on data availability, such that we were left with far fewer criteria incorporated into our metric, however these factors did still contribute to our final analysis, when determining the comparison of each candidate's habitability potential.

# 2. 1. 3. Polyextremophiles and the Limits of Life

The limits of life are determined by physical and geochemical parameters, within which biological functions such as metabolism must be sustained (Pikuta et al. 2009). Physical parameters include temperature, pressure, and radiation, while geochemical parameters include pH and salinity; their extremes within the biosphere occur in fringe environments and their boundary conditions are marked by polyextremophiles (Takai, 2019) given in Table 1. Organisms exist on bell–curve distributions, where maxima represent optimal conditions under which life proliferates, and minima represent the survival/tolerance limits of life (Merino et al. 2019).

**Table 1** Current habitability limits defined by known extremophiles for six selected crucial macroscopic factors collated from our literature review. These microorganisms define survivability thresholds as opposed to those of growth and maintenance

| Factor | Mes. | Limit / Defining Extremophile | Tol. | Pro. | Earth Extreme (Fringe Biosphere) |
|--------|------|-------------------------------|------|------|----------------------------------|

| Parameter | Range | Limit | Organism | | | Environment |
|---|---|---|---|---|---|---|
| Temperature (°C) | 15–45 | Upper Limit | *Methanopyrus kandleri 116* | 122 | 105 | 407 ℃ Mid–Atlantic Ridge Deep Hydrothermal Vent |
| | | Lower Limit | *Planococcus halocryophilus Or1* | -15 | 25 | -89 ℃ Vostok, Antarctica |
| Pressure (MPa) | 0.1 | Upper Limit | *Thermococcus piezophilus CDGS* | 125 | 50 | 110 MPa Challenger Deep, Mariana Trench |
| | | Lower Limit | *Bacillus subtilis* | 0.0025 | 0.1 | 0.0006 MPa Stratosphere at 56km |
| UV Radiation (W/m²) | - | Upper Limit | | - | - | 17.5 W/m² (UVB), 6.4 W/m² (UVC) ISS Thermosphere |
| Ionising Radiation (kGy/hr) | - | Upper Limit | *Thermococcus gammatolerans* | 30 | - | 1 kGy/hr Ancient 2Ga Deep Fissure Reactors |
| pH | 5–8 | Lower Limit | *Picrophilus oshimae KAW 2/2* | -0.06 | 0.7 | -3.6 Richmond Mine Pool Waters, Iron Mountain |
| | | Upper Limit | *Serpentinomonas sp. B1* | 12.5 | 11 | 13.1 Serpentine Seamount Pore Waters, Mariana Forearc |
| Salinity (%) | 0–5 | Upper Limit | *Halarsenatibacter silvermanii SLAS-1* | 35 | 35 | 35 % Saturated Soda Lakes (High–T) |

The diversity of life is not fully represented by the diversity of environmental niches and vice versa, as the conditions of life are not necessarily coterminous to environmental conditions on Earth (Cockell et al. 2016). Subsequently where life thrives under environmental maxima (e.g. macrofaunal presence), the limits of life become environment-limited and must be corroborated to extremophile cultures in experimental settings. Some limits, such as the Upper Temperature Limit (UTL) are well constrained, and despite the evolving knowledge of polyextremophiles (limited by discovery and the difficulty of experimental methods) are expected to fall within narrow error margins. Conversely, environment-limited settings such as the Upper Pressure Limit (UPL) may fall short of true limits that life can tolerate by several magnitudes (Takai, 2019).

# 2. 1. 4. Earth–Analogue Environments

The diversity of environments on Earth can be broadly split via the systems approach into the hydrosphere, atmosphere, and solid Earth; these systems are in constant exchange with life (biota) and with each other (Kump et al. 2010). We revise this categorization in accordance with Hays et al. 2015 in order to allow better analogue comparisons between Earth and other planetary bodies, splitting systems into surface, subsurface and hydrosphere environments. Surface and subsurface environments are further split by composition into icy vs. silicate environments to distinguish between ice crusts and polar ices from silicate crusts. Ocean environments are divided into ambient ocean, deep ocean floors and hydrothermal vent systems (Figure 1). This selection has been weighted in favour of water–rich environments, given that the presence of water is the prime criterion for life, and was informed by qualitative studies on inferred habitable environments on other planetary bodies (Preston and Dartnell, 2014; Hays et al. 2015; Merino et al. 2019). Additionally, with the exception of the ambient ocean, these environments represent most of the fringe conditions observed on Earth and the limits of life as defined by microbial extremophiles (Takai, 2019).

Seven environments have been chosen to represent the diversity of habitable environments on Earth and other planetary bodies: (1) surface deserts, (2) surface ices, (3) deep continental subsurface, (4) subsurface ices, (5) ambient oceans, (6) deep ocean floors, and (7) hydrothermal vent systems. Their full ranges are given in Table 2. These ranges are wide and often represent bimodal to multimodal conditions (e.g. hydrothermal vent acidity or alkalinity) and almost all ranges are strongly skewed from median values, which therefore, cannot be taken as representative of each environment (e.g. ocean temperatures have a mean of 2℃ and a median of 50℃). Subsequently specific environmental analogues were chosen for use in our metric, obtained from Preston and Dartnell's (2014) compilation of 30 analogues for habitability across the solar system. Their median values, used in our metric, are given in Table 3.

**Table 2** Full value ranges of our six environmental factors for all examples of our selected analogue environment types across Earth

| Analogue Environment | Temperature (C): | Salinity (% NaCL): | Pressure (bar) | Acidity/ Alkalinity (pH): | UV-c Radiation (W/m$^2$) | Radioactivity (Gy): |
|---|---|---|---|---|---|---|
| Icy Poles | -20–37.9 | 0.087–0.11 | -0.06–13 | 0 - 4 | 10–300 | 0.003–50 |
| Surface Continent | -98.6–24.3 | 0.11–35.5 | 4.6–9.6 | 0–40.2 | 0–551 | 0.3–1 |
| Subsurface Continent | 10–32.8 | 1–1504 | 6.55–8.46 | 0.0089 | 0 | 504–824 |
| Subsurface Ices | -1.8 - 0 | 1–10 | 4–8 | 3.5 | 0 | 1.5 |
| Ambient Ocean | -2–4 | 10–110 | 7.98–8.49 | 3.4 | 0 | 1.5 |
| Deep Ocean Floor | 2.1–112 | 7.3–8.1 | 3.4–3.9 | 1.9–13.8 | 0 | 3-100 |
| Hydrothermal Vents | 1–464 | 2.1–50.7 | 4–11 | 0.1–8 | 0 | 18.7-1074.5 |

**Table 3** Mean values for each environmental factor in each of the specific local environments we chose as examples of our seven generalised analogues

| Analogue Environment | Temperature (C): | Salinity (% NaCL): | Pressure (bar) | Acidity/ Alkalinity (pH): | UV-c Radiation (W/m$^2$) | Radioactivity (Gy): | Reference |
|---|---|---|---|---|---|---|---|
| Dry Valleys, Antarctica | -23 | 0.08 | 7.8 | 2.5 | 160 | 0.5 | Goordial et al. (2016); Wood et al. (2018); Finneran et al. (2013); Murphy (2020) |
| Atacama Desert | 19 | 0.11 | 8 | 15 | 280 | 50 | Crits-Christoph et al.  (2013); Voight et al. (2019); Cordero et al. (2018) |
| Columbia River Basalts | 70 | 40 | 9.3 | 0.35 | 0 | 504 | Stevens and McKinley (1995, 2000); Newcomb, (1972); Maxwell et al. (2013) |
| Lake Vostok, Antarctica | -3 | 35.5 | 5.6 | 0.08 | 0 | 1.5 | Siegert et al. (2001,2003); Lin et al. (2020) |
| Atlantic Ocean | 4 | 25 | 7.7 | 3.46 | 0 | 1.5 | Foustoukos,Savov and Janecky (2008); Murray (2004); Lin et al. (2020); Charmasson et al. (2009) |
| Mariana Trench | 1 | 110 | 7.8 | 3.48 | 0 | 3 | Preston and Dartnell (2014); Nunoura et al. (2015); Lin et al. (2020); Charmasson et al. (2009) |
| Lost City, Mid–Atlantic | 66 | 8.4 | 10 | 2 | 0 | 28.27 | Foustoukos, Savov and Janecky (2008); Preston and Dartnell (2014); Lang and Brazelton (2020); Lin et al. (2020); Charmasson et al. (2009) |

## 2. 2. Detailed Assessment of Main Targets

Our full list of targets gleaned from the literature is as follows; Mars, Europa, Enceladus, Venus, Ganymede, Titan, Io, Callisto, Ceres, Charon, Deimos, Phobos and Pluto. We now will look in detail at the targets that are most promising prospects for microbial life outside of Earth and have the most uniformly similar environments compared with other celestial objects in the solar system, henceforth known as our Main Targets. To do this we briefly take stock of all of our astrobiological targets gleaned from the literature and exclude a number from our primary group based on endogenic factors. Mercury, the planet nearest to the Sun, receives extreme radiation and is therefore unlikely to support life, followed by Venus, which has extreme temperatures and pressures (740K, ~95 bar), making its surface conditions inhospitable for life. It is notable that Venus has been the subject of recent astrobiological inquiry, but the nature of its geology and environments makes it unsuitable to compare with our other targets using our method, which are mostly icy-ocean moons. However, its atmosphere is quite similar to Earth's atmosphere and there have been speculations of it being hospitable to thermoacidophilic extremophiles (Dartnell, L.R. et al. 2015). The moon Charon and the dwarf planet Ceres, like our own moon, have a high radiation environment with the absence of proper sources of energy to support microbial life. Both have only a few water molecules embedded in the regolith and a large fraction of their cores is solidified, with endogenic sources of energy being scarce (Weber, R.C. et al. 2011). The two satellites of Mars; Deimos and Phobos, are less likely to be habitable because of their size, with no available source of energy and extremely less water content on their surfaces (~$10^{-2}$ %wt.) (Truong, N et al. 2016). Other targets like Jupiter's moon Io which have extreme radiation environments with eruptive volcanic mountains on the surface, are largely inhospitable. This leaves seven main targets, Mars, Europa, Enceladus, Titan, Ganymede, Callisto and Pluto which we explore individually in more detail in this section.

## 2. 2. 1. Mars as an Astrobiology Candidate

Mars is a hyper-arid, rocky and cold planet. It is the most popular target for the search of microbial life beyond Earth. It is believed that Mars once had liquid water bodies spanning its surface; lakes and rivers, and even an ocean in the northern hemisphere (Sharabian et al. 2020). However, over time, it lost a major part of its atmosphere, underwent dehydration and the remaining water turned into ice. Due to its thin atmosphere and lack of an active magnetic field, it receives many different forms of intense radiation like Solar Energetic Particles (SEPs), Galactic Cosmic Rays (GCRs) and Ultraviolet (UV) radiation. RAD (Radiation Assessment Detector) data from Mars Curiosity rover measured a GCR dose value of 76 mGy/yr and a GCR Dose Equivalent Rate of 232 mSv/yr (Hassler et al. 2014, Zeitlin et al. 2017) at Gale crater on the surface of Mars. The existence of extremely large volcanoes on the Martian surface indicates that, over time, the planet has been getting exhausted of its internal heat. Although bodies of stable water do not exist at present, except for transients at higher latitudes, the water-rich minerals composed of phyllosilicates and sulphates in the Mars regolith make it a potentially suitable place for microbial life. In 2012, Curiosity rover also obtained the evidence of an old streambed in the Gale crater and suggested vigorous flow of water in the past. Gale Crater, Columbia Hills (Gusev Crater), Eberswalde Crater, Jezero Crater, Mawrth Vallis, Nili Fossae, Holden Crater, NE Syrtis, SW Melas, Miyamoto Crater and Southern Meridiani all show similar features. These environments are also abundant in water-rich minerals; Olivine, silicates, Fe-oxides, carbonates, calcium Pyroxenes, and sodium and potassium rich feldspar. Apart from these, we have also included the Ultimi Scopuli - which was revealed by radar data to be a subglacial lake environment beneath the crust in the southern region, in our catalogue of Mars environments, though this is contested (Lauro et al. 2021). See Table 4 for collected data.

**Table 4** Mean values for each environmental factor in each of the specific local environments chosen for Mars

| Environment | Temperature (C): | Salinity (% NaCL): | Pressure (bar) | Acidity/ Alkalinity (pH): | UV-c Radiation (W/m²) | Radioactivity (Gy): | Reference |
|---|---|---|---|---|---|---|---|
| Icy Poles | -129.3 to -124 | 1.2-6 (3) | 0.00691-0.00705 | 9-12(10) | 74 - 87 | 0.83/yr | Forget et al. (1999); Merino et al. (2019); Zolotov et al. (2004); Dartnell et al. (2008) |
| Surface Desert | -89 to -46(-68) | 5.2 - 5.8 | 0.00691-0.00705 | 8.3 | 282 | 0.83/yr | Forget et al. (1999); Merino et al. (2019); Fairen et al. (2008); Dartnell et al. (2008) |
| Subsurface (Continental) | -33 to -23.15 | 0 | ~300 | 10-12(11) | 0 | 0 | Hoffman et al. (2001); Jones et al. (2011); Zolotov et al. (2004) |
| Subsurface Ice (Ultimi Scopuli) | -36.4 to -80(-68) | 10 | ~100 | 9-12(10) | 0 | 0 | Orosei et al. (2018); Zolotov et al. (2004); Jones et al. (2011) |

## 2. 2. 2.  Europa as an Astrobiology Candidate

Europa is an icy moon in the Jovian system composed of an icy shell, a global subsurface ocean and a rocky core. The icy shell is thought to be several kms thick and the subsurface ocean 100km thick (Lipps et al. 2005). Europa has the youngest (least cratered) surface in our solar system, suggesting recent–active geological processes (resurfacing). Habitat divisions include craters, fissures/cracks, different layers of a stratified ocean, hydrothermal vents and soft substrata on the ocean floor (Lipps et al. 2005). Life on Europa would be subject to extreme conditions (low temperatures, high pressures, high surface radiation and variable ocean salinity, low pH). Radiolytic oxidant production on the surface (which resulted in the tenuous O2 atmosphere) can enrich material at depth: these include O2, H2O2, CO2, CO and other organics (e.g. carbon from exogenous delivery) (Hand et al. 2007). The subsurface ocean is least likely to be habitable in the majority of its volume – chemical gradients will be highest where nutrient upwelling (suspension) and downwelling (leakage) is highest (Lipps et al. 2005). For radiolytically produced oxidants to accumulate, we must compensate for hydrothermal reductants. Delivery timescales are plausibly less than 100Ma even if they are the same as Europa's surface age of 30–70Ma (Zahnle et al. 2003; Hand et al. 2007). If delivery timescales are adequately low, dissolved oxygen in Europa's oceans will be sufficient to support Earth-like macrofauna as found in the oxygen minimum zone. Europa's colder water and higher pressures mean oxygen saturation would be much higher than Earth's (50mM at –3℃ and 10–100MPa). These 'benthic' habitats include hard substrates, soft sediments, and hydrothermal vents (Lipps et al. 2005). Vents on Earth provide their own energy sources, have high chemical and temperature gradients, and support chemosynthetic bacteria (chemoautotrophs) as well as a host of protozoans and metazoans. Biota in Europan analogues would be expected to be chemoautotrophic, with rapid growth, fast metabolism, and high biomass. See Table 5 for collected data.

**Table 5** Mean values for each environmental factor in each of the specific local environments chosen for Europa

| Environment | Temperature (C): | Salinity (% NaCL): | Pressure (bar) | Acidity/ Alkalinity (pH): | UV-c Radiation (W/m$^2$) | Radioactivity (Gy): | Reference |
|---|---|---|---|---|---|---|---|
| Icy Poles | -177 | 0-10 (5.6) | 1.0E-13 | - | 0.03 W/m2 | 0.3Gy/yr | Ashkenazy et al. (2019); Zolotov et al. (2001); McGrath et al. (2009); Melosh et al. (2004) |
| Surface Desert/ Plains | -173 | 0 | 1.0E-13 | - | 0.05 W/m2 | 1971 /yr | Melosh et al. (2004); McGrath et al. (2009); Teodoro et al. (2016); |
| Base of Ice crust | 0 | 3 | 20 | 2.3 | 0 | 0.3Gy/yr | Melosh et al. (2004) |
| Ocean/ Deep Ocean | 4 | 3 | 20–130 | <8.5 or >8.5 | 0 | - | Melosh et al., (2004); Naganuma et al. (1998); Johnson et al. (2019) |
| Ocean Floor | 4.5 | 4 | 130–260 | <8.5 or >8.5 | 0 | - | Hand et al. (2007); Naganuma et al. (1998); Johnson et al. (2019) |
| Hydrothermal vents | 365 | 4 | 130–260 | <8.5 or >8.6 | 0 | - | Takai et al. (2008); Naganuma et al. (1998) |

## 2. 2. 3. Enceladus as an Astrobiology Candidate

Enceladus, the second outermost moon of Saturn, is rich in water and several organics which made it one of the primary targets in search of microbial life. During the flybys of Cassini-Huygens mission, the Cassini spacecraft made the significant discovery of finding the geysers that spew from the SPT (South Polar Terrain), the location of which contains the geologically active "Tiger stripes" (Glein et al. 2018) and also made measurements of the gaseous composition of the plume and detected the presence of methane, ethane, water, oxygen and carbon amongst other organics embedded in the surface ice (Parkinson et al. 2007). The Visible and Infrared Mapping Spectrometer (VIMS) determined that the surface of Enceladus is nearly pure water-ice, with the exception of the organics (Brown et al. 2006). The spewing of particles and chemistries in the plumes indicates a global subsurface ocean beneath the crust with active hydrothermal vents. Decay of short-lived radionuclides such as $^{238}$U and $^{40}$K, as well as tidal heating, causes the melting of water-ice, which leads to water-rock interactions in the seafloor, with subsequent cycles of aqueous alteration of the rocky core changing oceanic composition (Zolotov et al. 2007). With such a high water to rock ratio, Enceladus is less salty and has less inorganic carbon than terrestrial seawater, and an alkaline pH value of 9 – 11 (Glein et al. 2016). Active hydrological cycles, hydrothermal vents and cryovolcanoes help to transfer the organic materials to the surface and could potentially serve as an energy source for microbial life (Tsou et al. 2012). See Table 6 for collected data.

**Table 6** Mean values for each environmental factor in each of the specific local environments chosen for Enceladus

| Environment | Temperature (C): | Salinity (% NaCL): | Pressure (bar) | Acidity/ Alkalinity (pH): | UV-c Radiation (W/m$^2$) | Radioactivity (Gy): | Reference |
|---|---|---|---|---|---|---|---|
| Icy Poles | -159 – -116 | 0 | 0 | 0 | 7.52*10^-6 | 0.3/yr | Porco et al. (2006); Parkinson et al. (2007); Teodoro et al. (2017) |
| Surface Desert/ Plains | -201 | >0.5 | 0 | 8.5–9 | 1.1 × 10^12 | 0.3/yr | Hsu et al. (2015); Parkinson et al. (2007); Teodoro et al. (2017) |
| Base of Ice crust (Subsurface) | 0 | 0-4(4) | 1-2(1) | 8.5–10.5 | 0 | 0 | Hsu et al. (2015); Glein et al. (2016) |
| Ocean/Deep Ocean | 3 | 0.5-2 | 1–8 | 11-12 | 0 | 0.05 | Glein et al. (2018); Hsu et al. (2015); Teodoro et al. (2017) |
| Ocean Floor | 87 | 0-4(4) | ~7.4 | >8.5 | 0 | 0 | Glein et al. (2018); Hsu et al. (2015); McKinnon et al. (2015); |
| Hydrothermal vents | >90 | 3 - 10 (4) | 7.4 - 37 | >8.5 | 0 | 0 | Hsu et al. (2015); McKinnon et al. (2015); Glein et al. (2018) |

# 2. 2. 4. Titan as an Astrobiology Candidate

Titan, the largest moon of Saturn and the second largest in the solar system, has always been a unique astronomical object. The presence of a thick nitrogen, methane and hydrogen atmosphere gives it a hospitable pressure of 1.45 atmospheres (Jennings et al. 2016), and a special relevance for astrobiological purposes amongst other moons (Catling et al. 2017). It is now believed that Titan is home to many essential precursors for the formation of life, as we believe were present in the prebiotic epoch on our own planet (Raulin, 2005; Gudipati et al. 2013). Organic chemistry is also highly likely to be active in Titan's atmosphere, driven by solar radiation and Saturn's magnetic winds. Titan is home to gargantuan surface lakes of hydrocarbons, mostly methane and ethane, as well as a potential liquid water subsurface ocean (Grasset et al. 2000). Erosion features and river deltas, and many earth-like fluvial features are attributed to an active methane/ethane cycle analogous to the water cycle on Earth, with dunes, valleys and tributaries carved by dark hydrocarbon deposits and rain (Lorenz, 2003; Lorenz et al. 2006). However, requirements of the methanogenesis process in the surface lakes and the necessarily incredibly high salinity of the subsurface oceans give any potential life a particularly difficult struggle to overcome (Fortes, 2000). Possible photochemical dissociation of ammonia could be responsible for the generation of tholins, as shown by Miller-Urey (Hill et al. 2003; Raulin et al. 2012) but a biological source has not been ruled out. The organic gases of ethane, propane, acetylene, ethylene, hydrogen cyanide and cyanogen were also observed in trace amounts. Surface spectra measurements which are difficult to recreate in the lab using only the ices, organics and tholins as well as the apparent replenishing and seasonal variation of methane, suggest exchange of organic material and water ice from beneath the surface (Niemann, 2005), and a water-ammonia subsurface sea could be feeding material to the surface through cryovolcanoes and other means (Mousis and Schmitt, 2008; Mitri, 2014). This would also explain other observed anomalous features, including orbital eccentricity and magnetic field discrepancy (Perkins, 2012; Mitri, 2014). All these factors point to Titan as an incredibly valuable astrobiological target.  See Table 7 for collected data.

**Table 7** Mean values for each environmental factor in each of the specific local environments chosen for Titan

| Environment | Temperature (C): | Salinity (% NaCL): | Pressure (bar) | Acidity/ Alkalinity (pH): | UV-c Radiation ($W/m^2$) | Radioactivity (Gy): | Reference |
|---|---|---|---|---|---|---|---|
| Icy Poles | -184 | - | 1.5 | - | 0 - 0.35 W/m2 | - | Mousis et al. (2008); Stevenson et al. (1986) |
| Surface Desert | >-183, <-179 (-181) | - | 1.5-3.5 (2.5) | - | 0 - 0.35 (W/m2) | - | Merino et al. 2019); Stevenson et al. (1986) |
| Base of Ice crust | -18 | - | 500-3000 | 11.8 | 0 | - | Merino et al. (2019); |

| Ocean/Deep Ocean | - | 30 - 40 (35) | - | - | 0 | - | Mousis et al. (2008) |
| Ocean Floor | - | - | - | - | 0 | - | |
| Hydrothermal vents | - | - | - | - | 0 | - | |

## 2. 2. 5. Ganymede as an Astrobiology Candidate

Ganymede, the largest satellite in our solar system, is an icy ocean world. Its surface is predominantly covered by water ice (50-90% by mass) and it is believed to have a complex and turbulent geological history. It is the only satellite in the solar system that has an intrinsic magnetic field (Showman and Malhotra, 1999). Measurements of Ganymede's axial moment of inertia reveal the possible existence of a subsurface liquid water ocean, estimated at a depth of 150km (Carlson et al. 1973). An explanation for this water in liquid form is the heat generated through tidal dissipation and radiogenic energy (Grasset et al. 2013). Data collected by Voyager and Galileo show that Ganymede satisfies some of the prerequisite conditions to harbour life, for example, aided by internal pressure, polymerization can be induced, thereby affecting the reactivity of simple organic molecules (Pappalardo et al. 1998). This in turn, can enhance and support catalytic reactions and play a role in the stability and conformation of various biomolecules. The subsurface ocean may be inhabited by extremely barophilic organisms like endospores of proteolytic type B Clostridium botulinum TMW 2.357 (Grasset et al. 2013). Owing to the high radiation surface environment on Ganymede, organisms cannot survive on it for a prolonged period. Nevertheless, there is a possibility of endolithic organisms penetrating the subsurface, repairing the cellular damage due to radiation through slow metabolism and surviving by burying themselves deep in the subsurface (de Kleer et al. 2021). Considering all the above possibilities, Ganymede is an interesting target for the existence of organics and microbial life, or even a spacecraft-mediated biological transfer and propagation of life.  See Table 8 for collected data.

**Table 8** Mean values for each environmental factor in each of the specific local environments chosen for Ganymede

| Environment | Temperature (C): | Salinity (% NaCL): | Pressure (bar) | Acidity/ Alkalinity (pH): | UV-c Radiation (W/m$^2$) | Radioactivity (Gy): | Reference |
|---|---|---|---|---|---|---|---|

| Icy Poles | >-161, <-158 | - | >0.2E-11, <1.2E-11 | - | - | - | de Kleer et al. (2021); Hall et al. (1998) |
|---|---|---|---|---|---|---|---|
| Surface Desert/Plains | >-183, <-121 (-153) | - | >0.2E-11, <1.2E-11 | - | - | 18.25-29.2 Sv/yr | Spohn and Schubert (2003); Hall et al. (1998); Podzolko and Getselev (2013) |
| Subsurface | -73 to -43 (-61) | - | 1.00E+13 | 9.3 | 0 | - | Freeman (2006) |
| Base of Ice crust | -43 to -3 (-21) | - | 1.00E+20 | 9.3 | 0 | - | Freeman (2006) |
| Ocean/Deep Ocean | -3 to -21 (-18) | 3-10(5) | 1.00E+20 | 9.3 | 0 | - | Vance et al. (2014) |
| Ocean Floor | 7 | - | 1.00E+22 | ND | 0 | - | Vance et al. (2014) |
| Hydrothermal vents | | - | 1.00E+08 | | 0 | - | |

## 2. 2. 6.  Callisto as an Astrobiology Candidate

Callisto is characterised by unique surface processes, landforms and evolution. Callisto was initially perceived as a dead moon, considering the absence of surface endogenic activity. $SO_2$ is most probably present in the form of molecules trapped in the surface and subsurface materials of Callisto. It is thought to be primordial or evolved from endogenic processes on the satellite (Moore et al. 2004). Callisto has a global atmosphere composed primarily of $CO_2$ and $O_2$ (Carlson, 1999). The trend of $SO_2$ distribution on Callisto is opposite to that of $CO_2$ and the distribution is asymmetric. $CO_2$ is present in the recently formed craters as well, which implies that it gradually sublimes from the bedrock (Anderson et al. 1999). It could also be from delivery by comets, impact generation and indirect sublimation (Hartkorn et al. 2017). The atmosphere may also contain gaseous $H_2O$ since water-ice covers a significant portion of Callisto's surface. Considering its stagnant surface, Callisto is thought to have an ocean that is almost completely frozen (Vance et al. 2018), but might be interacting with rocks, thereby increasing the possibility of the existence of microbial life. However, whether such an ocean could provide a suitable habitat to resilient forms of life is still being debated (Bender et al. 1997). See Table 9 for collected data.

**Table 9** Mean values for each environmental factor in each of the specific local environments chosen for Callisto

| Environment | Temperature (C): | Salinity (% NaCL): | Pressure (bar) | Acidity/ Alkalinity (pH): | UV-c Radiation (W/m$^2$) | Radioactivity (Gy): | Reference |
|---|---|---|---|---|---|---|---|
| Surface Desert/ Plains | >-193, <-108 (-139.2) | - | 7.5 | - | 0.005263 W/m2 | 0.0365 Sv/yr | Moore et al. (2004); Carlson et al. (1999); Freeman (2005); Fredrick (2000) |
| Subsurface | -33 to 42(2.25) | - | 10^7 | - | 0 | - | Freeman (2005); |
| Base of Ice crust | -13 to 62(19) | - | 10^13 | - | 0 | - | Freeman (2005) |
| Ocean/Deep Ocean | -3 to 103(38) | - | 10^14 | - | 0 | - | Freeman (2005) |
| Ocean Floor | 2 to 127(50) | - | 10^16 | - | 0 | - | Freeman (2005) |

# 2. 2. 7. Pluto as an Astrobiology Candidate

The lonely dwarf planet Pluto essentially has a rocky core encapsulated by a mantle of water-ice, with a surface covered by a variety of ices like methane and nitrogen frost (Stern et al. 2015). Pluto has a complex geology, the details of which were revealed by NASA's New Horizons spacecraft. Ongoing surface geological activity was observed, including regions having active glacial flow, movement of massive water-ice blocks and pitting. The variety of terrains on Pluto include Sputnik Planum (an area of ~870,000 km$^2$ comprising $N_2$, CO and $CH_4$), rugged mountainous regions and many other heterogeneous surfaces with varying crater densities and surface textures. Along the western margin of Sputnik Planum lies a range of mountains with seemingly random orientations and altitudes of up to 5 km (Moore et al. 2016). The existence of methane ice on Pluto indicates that it probably has an atmosphere consisting of heavy gases, the absence of which, would result in quick sublimation of the methane ice. Pluto's atmosphere contains extensive, distinct, and optically thin layers of hazes. Molecular nitrogen ($N_2$) is dominant, whereas methane ($CH_4$), ethane ($C_2H_6$), acetylene ($C_2H_2$) and ethylene ($C_2H_4$) are present in lesser quantities. A secondary source of methane photolysis is provided by the interplanetary hydrogen scattering of solar Lyman alpha photons. Also, high-altitude haze nuclei are produced by ionisation of $N_2$ leading to the formation of large ions (Gladstone et al. 2016). Considering Pluto's extremely cold surface, it is unlikely to be habitable for life as we know it. However, its warmer interiors might be a better candidate. See Table 10 for collected data.

**Table 10** Mean values for each environmental factor in each of the specific local environments chosen for Pluto

| Environment | Temperature (C): | Salinity (% NaCL): | Pressure (bar) | Acidity/ Alkalinity (pH): | UV-c Radiation (W/m$^2$) | Radioactivity (Gy): | Reference |
|---|---|---|---|---|---|---|---|
| Icy Poles | -387 to -369 F | - | 1.30E-05 | - | - | - | Cruikshank et al., (1980); Stern et al. (2015) |
| Surface Desert/Plains | >-240, <-218 (-229) | - | 1.30E-05 | | 0.873 | 0 | Cruikshank et al. (1980); Stern et al. (2015) |
| Subsurface | - | - | - | - | 0 | | |
| Base of Ice crust | - | - | - | - | 0 | | |
| Ocean/Deep Ocean | - | - | - | - | 0 | | |

# 3. Methodology

## 3. 1. Overview

Our method essentially seeks to quantify, gauge and compare the habitability of worlds in our solar system, with reference to the survivability criteria of known species of microbes here on Earth. We seek to do this by focusing on the limits of extremophiles, and the boundary values in extreme biospheres on Earth where extremophiles have been known to be capable of surviving. After collecting data on the appropriate survivability factors to consider, then cross-referencing with the environmental data we have for the targets under consideration (which have already been filtered by bulk data and endogenic factors), we must reduce the number of criteria incorporated to account for data availability. This leaves us with seven environments /biospheres: Icy Poles, Continental Surface, Continental Subsurface, Ice Subsurface, Oceans, Sea Floor and Hydrothermal Vents; and six microbial habitability factors: Temperature, Pressure, Salinity (%NaCl), Acidity (pH), UV-c Radiation and Radioactivity, which we will analyse across each environment and target. Of course, there are also other extrinsic environmental factors which have some bearing on habitability - namely cosmic dangers and gradual environmental changes over a long timescale - which cannot be quantified numerically. We take up discussion of how these factors could affect our results in the conclusion.

Our method for analysis is conducted in two stages. First, we identify which of our targets have environmental variable values falling within the limits of the microbial habitability factors, also assessing the most prevalent factors and which types of environments consistently fall into habitability ranges across our targets. In the second stage we implement a new method, the so-called Microbial Habitability Index (MHI) scoring system, to determine how close each of our target environments is - in terms of environmental factor values - to the factor values measured in analogue environments, which we know to be habitable and hospitable to microbes here on Earth. The method for the second, more thorough stage of analysis, is laid out next.

# 3. 2. 1.  MHI Scoring System

Our method uses as its foundation a modified version of the Earth Similarity Index (ESI) to give independent environment scores between 0 and 1, a score of 1 indicating perfect agreement of the target environment with the corresponding Earth analogue environment across every environmental factor. We then sum across these individual environment scores to give an overall habitability score, normalised by the number of environments. This data is then used to rank the habitability of our primary candidates, both on an individual environmental and overall habitability basis.

To compute the environmental habitability index (EHI), we define the following metric:

$$EHI = \prod_{i=1}^{n} \left( 1 - \left| \frac{x_i - x_{i0}}{x_i + x_{i0}} \right| \right)^{\frac{1}{n}}$$

(1)

Here the $x_i$ are the values for microbial survival factors for the particular target environment, the $x_{i0}$ are the same factors in the corresponding Earth analogue environment and we multiply across all $n$ factors in series to compute the EHI for the chosen target and environment. The next step in the process is to calculate an EHI for each of our m environment types, then sum and normalise to give an overall MHI score between 0 and 1. Here an ideal value of 1 would imply every environment of the target matches the habitability of the Earth analogue in every selected vital requirement for microbial survival exactly, whereas a zero would imply it is absolutely inhospitable. The total formula for the MHI for a given target is then summarised as:

$$MHI = \frac{1}{m} \sum_{j=1}^{m} EHI_j$$

(2)

Where EHI is defined as before in equation (1), $j$ is the index of our environment types, and $m$ is the total number of environments. In this work we have considered $m$ as 7 and $n$ as 6, but the approach is scalable for any number of environments that can be identified on a given habitability target, provided there is sufficient data for each of the microbial survival criteria factors for each of those environments. Expanding the definition of EHI we can alternatively re-write the MHI definition (where the index notation preserves its meaning and $x_{ij}$ is the i$^{th}$ factor of the j$^{th}$ environment) as:

$$MHI = \frac{1}{m} \sum_{j=1}^{m} \left[ \prod_{i=1}^{n} \left( 1 - \left| \frac{x_{ij} - x_{ij0}}{x_{ij} + x_{ij0}} \right| \right) \right]^{\frac{1}{n}}$$

(3)

In 3.2.3. we give a hypothetical example to show how to use this method in practice. In Table 1 we have the range and mean of the habitability factor values, represented as: lower bound – upper bound (mean) for an imaginary target, in each of the seven environment categories. Table 2 has the same information entered its cells, but with respect to Earth analogue environment reference values, the same as we used for our actual analysis.

## 3. 2. 2. Caveats

There are a few shortcomings of the method which affect the validity of results and are worth addressing now. The first point regards the metric definition itself, which needs to be amended for certain fringe cases, for example in the case of a zero-value cell, or a case where a positive value is being compared to a negative value for a particular environment. In the former, a value of 0 for a particular factor, as in the case of UV-c radiation in subsurface environments, will give a difference and sum both equivalent to the other non-zero value being compared against, hence their ratio will be 1 and the metric will become zero for that term in the multiplicative series, of course making the whole expression zero and giving a null value for the habitability score. To amend this, we set all of our zero values to 1 x 10⁻⁶, three orders of magnitude less than our three significant figure final score. Additionally, in the case where a negative value is compared with a positive value, this has the possibility of the difference being greater than the sum, giving a negative metric value. To remove this possibility, we add the absolute value of the negative criterion value plus 1 x 10⁻⁶ to both amounts being compared, such that the target value is the same relative amount away from the reference ideal value, but we get a positive metric score less than one as required for the multiplicative series to give us an overall positive value less than one.

# 3. 2. 3. Exoplanet Hypothetical Example

To demonstrate how this metric system is implemented in practice, we take the hypothetical example of an exoplanet, which we will call EXO1, and attempt to find EHI and MHI scores with our method. Let's propose the exo-planet is known to have a thin atmosphere, icy poles and a continental and icy subsurface, with known factor data in each environment. EXO1 is believed to have an ocean as well due to orbital anomalies and density calculations, but no specific data is known for any of our factors in this subterranean regime. This means our first step is to exclude our water environments from the MHI calculation, e.g Ocean, Sea Floor and Hydrothermal vents. Below is a table (Table 11) with hypothetical values for the factors in the respective environments of EXO1.

**Table 11** Habitability values for hypothetical exoplanet EXO1, and we suppose there is no available data for three environment types: Ocean, Sea Floor and Hydrothermal vents

| Environment | Temperature (C): | Salinity (% NaCL): | Pressure (bar): | Acidity (pH): | UV-c irradiation (W/m2): | Radioactivity (Gy/yr): |
|---|---|---|---|---|---|---|
| Icy Poles | -119.3 to -134 (-120) | 0.2 to 4 (2) | 0.5 to 0.7 (0.6) | 7 to 10(8) | 44 to 57(50) | 0.5 to 0.55 (0.53) |
| Continental Surface | -99 to -56(-68) | 3.2 to 3.8(3.5) | 0.06 to 0.08 (0.007) | - | 214 to 240 (232) | 0.4 to 0.6 (0.53) |
| Subsurface Continental | -23 to - 13.15 (-20) | - | - | 9 to 11(10) | 0 | 0 |
| Subsurface Ice | -26.4 to -70(-78) | 5 to 12(10) | 42 to 61 (50) | 9 to 12(9) | 0 | 0 |
| Ocean | - | - | - | - | - | - |
| Ocean Floor | - | - | - | - | - | - |
| Hydrothermal Vents | - | - | - | - | - | - |

First of all, note that we have left the final three rows blank, as Ocean environments are not confirmed for EXO1. Additionally, of the environments that are known, there is missing data for some of the factors, which we are supposing in this hypothetical case, have not been modelled or estimated. This is to replicate the kind of data availability our team was presented with when applying this method to astrobiological targets of interest in our solar system.

To calculate the MHI we need to take the values from appropriate cells and insert them into our metric equation to compute the summed score. As an example, let us suppose we wish to compute the EHI score for the continental surface environment. To do this we will be comparing each cell in row 3 column-wise with each corresponding cell in the Earth data values table. For the first cell we see a temperature of -68 degrees Celsius, to be compared with the temperature of 19 degrees in our Earth analogue environment, the Atacama desert. These two values go into our equation (1) as $x_i$ and $x_{i0}$ respectively. We get a score of 0.1181 for our temperature factor. Applying the same method to our remaining factors and multiplying across the row we get a total of 0.0479 for our EHI. Note our fourth cell is empty, so we do not include this in our value for $n$ in the metric equation or the multiplicative sum, i.e. our $n$ value for this row is 5 and for the Subsurface Continental environment our $n$ is 4. If we find an EHI for every applicable row (the first four) and take the mean, we get the MHI. For EXO1 we find this to be 0.1313.

# 4. Results

## 4.1. Boundary Values

The first stage of our analysis of this data involved checking how many selected environments in our targets fell within ranges of microbial survival as determined by extremophiles in particularly fringe biospheres. Below is a table (Table 12) specifying the number of factors that each target had for each target environment within ranges determined by the limits of the corresponding analogue environments in each factor.

**Table 12** Number of environments for each target with habitability factors in the range of the limits defined by extremophiles

| Environment | Mars | Enceladus | Europa | Titan | Callisto | Ganymede | Pluto |
|---|---|---|---|---|---|---|---|
| Icy Poles | 4/6 | 2/6 | 2/5 | 2/3 | - | 0/2 | 0/2 |
| Surface | 3/6 | 3/6 | 1/5 | 0/3 | 2/4 | 1/3 | 0/4 |
| Subsurface Continental | 1/6 | - | - | - | 1/3 | 1/4 | - |
| Subsurface Ice | 2/6 | 3/6 | 2/6 | 1/4 | 1/3 | 1/4 | - |
| Ocean | - | 3/6 | 2/5 | - | 1/3 | 1/5 | - |
| Ocean Floor | - | 3/6 | 2/5 | 0/2 | 1/3 | 1/3 | - |
| Hydrothermal Vents | - | 5/5 | 4/5 | - | - | 1/2 | - |

There are a total of six environmental factors, this is reduced in several cases when there is missing data. For example, in the case of the Icy Poles of Ganymede, where the literature contains examples of Temperature and Pressure data but no Salinity, Alkalinity, Radioactivity or UV Radiation data. Of the two factors with data in this environment, only the Temperature is within the limits of microbial habitability (the pressure is too low for any extremophile) hence the ½ value quoted in the appropriate cell entry. A blank cell signifies a null reading, where there is no such environment known or there is no data for a particular target and environment, e.g. Mars Ocean environments. We can see that from this data, Enceledus has the most environmental factors in range with 19, followed closely by Europa with 13. The smallest number of in-range environments belongs to Pluto, with none. In terms of fraction of in-range environments out of viable environments, Enceladus is still the most hospitable with 19/35.

We were also interested in finding which of our factors were the easiest to satisfy, and if there were any patterns across environments and targets. The following table represents the number of environments of each target which fall within limits for each of our microbial vitality factors. This time the numerator of the fractions represents the number of environments which satisfy our extremophile limit conditions, out of the number of environment types present for a given target (See Table 13).

**Table 13** Number of environments for each target with habitability factors in the range of the limits defined by extremophiles

| Factor | Mars | Enceladus | Europa | Titan | Callisto | Ganymede | Pluto |
|---|---|---|---|---|---|---|---|
| Temperature | 0/4 | 3/6 | 2/6 | 0/3 | 1/5 | 1/6 | 0/2 |
| Salinity | 1/4 | 5/6 | 4/6 | 0/1 | - | 1/1 | - |
| Pressure | 1/4 | 2/6 | 1/6 | 1/3 | 0/4 | 0/7 | 0/2 |
| Acidity | 1/4 | 2/6 | 1/4 | 0/1 | 1/1 | 0/3 | - |
| Radiation | 4/4 | 4/6 | 4/6 | 2/4 | 3/5 | 4/5 | 0/1 |
| Radioactivity | 3/4 | 3/5 | 1/3 | - | 1/1 | 1/1 | 0/1 |

From this data we can see again that Enceladus consistently had the most environments in the Earth analogue regime for each microbial survival factor. This is followed by Europa and Mars. As for factors, Salinity fared the best, in range for eleven environments, and notably five out of six and four out of six of Enceladus environments, which implies that Earth Halophiles are very likely to survive in their oceans, and radioactivity resistant extremophiles could certainly survive in and around hydrothermal vents on Enceladus if they are indeed present. It is notable that the factors of Temperature had seven environments in range, whereas Acidity and pressure were in range for five environments each. UV-c had the most in range overall with twenty-one environments in range, but we must remember this is because UV-c in the subsurface environments is zero, which will always be in the limits of microbial survivability. It is not particularly hard to satisfy this requirement (every target does) other than on surface environments. Of the factors that are in range we can plot scatter diagrams (Figure 1, Figure 2) in multiple dimensions to get a better grasp on spatial distribution, and any commonality in which factors prove easier to satisfy. In terms of the environments, we found a four-way tie between the surface, subsurface ice, icy poles and hydrothermal vents, for the most total environmental factors in range (10 each), followed by ocean and sea floor with 7 each. The most inhospitable environment was the subsurface continental environment which only had 3 factors in range across all our targets.

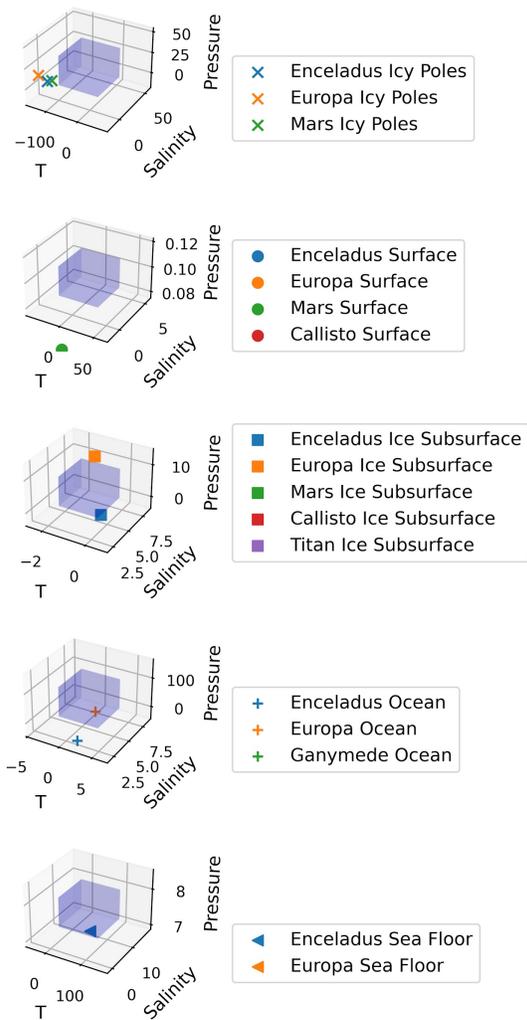

**Fig. 1** Scatter plots demonstrating where our environmental factors fall for each of our targets. The blue volume represents the boundaries of Earth life given by tolerance of known extremophiles for Temperature, Pressure and Salinity. An environment associated scatter point present inside this volume indicates that all factors displayed on the axis are in tolerable ranges.

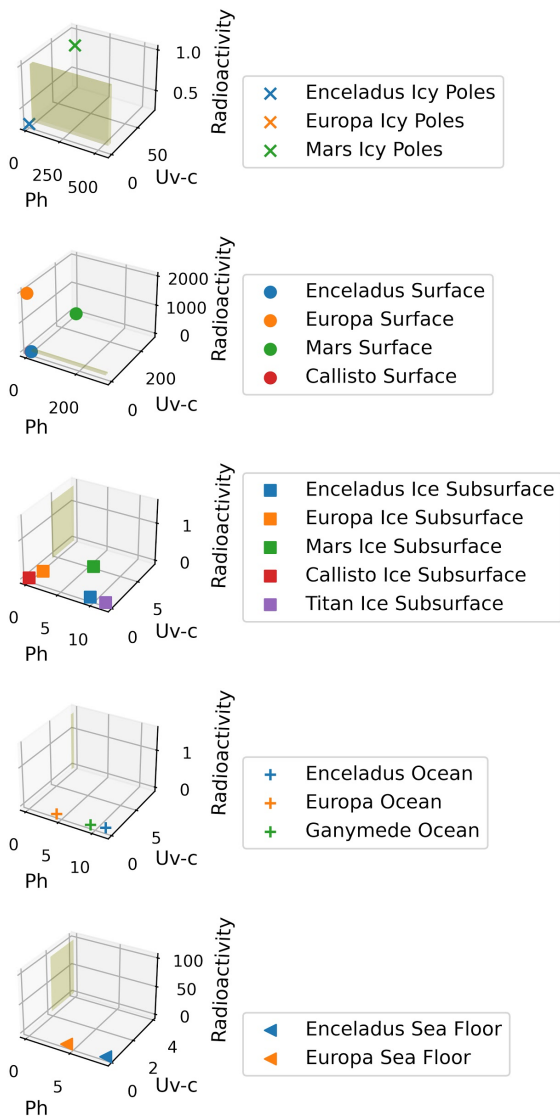

**Fig. 2** Additional scatter plots demonstrating where our environmental factors fall for each of our targets. The yellow volumes represent the boundaries of Earth life given by tolerance of known extremophiles for Radioactivity, Uv-c and Ph. An environment associated scatter point present inside this volume indicates that all factors displayed on the axis are in tolerable ranges.

## 4.2. Metric Score

Our next step was to use a combination of our EHI metric scores and overall MHI score to motivate a ranking of our considered targets in order of priority for future exploratory missions. Below we provide a table (Table 14) of our environment EHI scores and overall MHI scores for each target.

**Table 14. Microbial Habitability Scores for each of our environments across our targets, blank cells indicate cases where there was not enough data.**

| Target | Icy Poles | Surface | Continental Subsurface | Ice Subsurface | Deep Ocean | Sea Floor | Hydro-thermal Vents | Overall Score |
|--------|-----------|---------|------------------------|----------------|------------|-----------|---------------------|---------------|
| Mars | 0.4050 | 0.0793 | 0.0179 | 0.0348 | - | - | - | 0.1342 |
| Encela dus | 0.0055 | 0.0003 | - | 0.0178 | 0.4393 | 0.1039 | 0.7778 | 0.2241 |
| Europa | 0.0005 | 0.0000 | - | 0.0000 | 0.8013 | 0.7034 | 0.3986 | 0.3173 |
| Titan | 0.0339 | 0.0011 | - | 0.3351 | 0.4241 | - | - | 0.1986 |
| Ganym ede | 0.0000 | 0.0000 | 0.0001 | 0.0000 | 0.0000 | 0.0000 | 0.0000 | 0.0000 |
| Callisto | - | 0.0016 | 0.0037 | 0.0000 | 0.0000 | 0.0000 | - | 0.0011 |
| Pluto | 0.0066 | 0.0003 | - | - | - | - | - | 0.0034 |

As expected, we saw much higher scores for the water environments than for surface environments for icy ocean worlds like Enceladus and Europa. The continental surface surprisingly trumped the subsurface despite the influence of radiation and temperature factors, and icy polar regions consistently attained the highest scores amongst the non-subterranean environment categories. Below is a figure (Figure 3) of the scatter arrangement of scores against the environment, as well as trend lines for the average scores of Icy Ocean Moons and others.

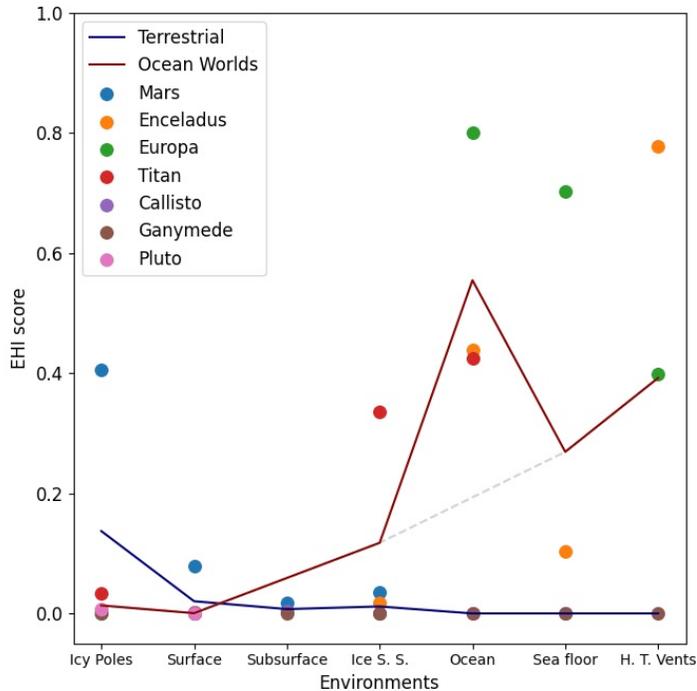

**Fig. 3** Environmental Habitability Index (EHI) scores for each of our considered targets represented by scatter points. The thicker lines show the trends of average EHI with environment type for terrestrial (i.e Mars, Ganymede, Callisto, Pluto) and Ocean World targets (Europa, Enceladus, Titan). Each subsequent environment along the x-axis is deeper below the surface. S.S. is shorthand for Subsurface, the 'Subsurface' environment is the continental subsurface.

We found that Europa had the highest overall habitability score of 0.3173, followed by Enceladus in second place with 0.2241, and Titan in third with 0.1986. The single highest individual EHI was of Europa's deep ocean environment, followed closely by the sea floor of Enceladus. The lowest EHI goes to Callisto's Icy surface Poles. It is clear for our Icy Ocean worlds, below the surface we have much better prospects for habitability. As the depth increases, so does the habitability score, with a peak in the ocean environments. The terrestrial targets, though subtler show the opposite trend, however we may note this is largely due to the absence of known water environments on e.g. Mars.

From this evidence alone we would conclude that the best environment to look for life in our solar system currently is Europa's deep ocean, followed by that of Enceladus. In the short term, Mars still looks like the best candidate to explore based on its MHI score and considering the current number of proposed missions, negotiable terrain and its relative proximity. If we also make considerations of energy availability, bio-essentials, and other factors (See 2.1.1), searching the deep oceans of Europa including the sea floor in search for hydrothermal vents would be our top suggestion for future astrobiological mission aims in the medium to long term. Enceladus and Titan are also very good overall candidates, and with the diversity in the latter's climate and geology we conclude they too should be prioritised for exploration.

# 5. Evaluation and Discussion

Intriguing comparisons can be made between certain surface conditions of our targets. Most of our surface environments are on bodies with no atmosphere, or a very tenuous atmosphere, and as such have extremely low (on the order of $10^{-6}$) scores, a result of incredibly low temperatures and pressures far outside the range of survivability of almost all known extremophiles and microbes. On the other hand, it is intriguing that we find such a large difference in the surface environments of Mars and Titan. Of course, these results have been hindered by the lack of specific environmental data for a lot of these targets. For instance, data of the Martian subsurface environment which was once habitable and may have had hydrothermal systems in the past, was missing; while with Titan, its subsurface ocean and ocean floor data were missing, which are rich in organic chemistries and habitable for microbial life. Venus has recently had the spotlight thrown on it, with possible organics present in the atmosphere, but not much can be said for the inhospitable surface below and as such it is excluded from our main targets, with no quantitative data available to use in a metric such as ours, until probes can visit Venus, which is still some way off despite NASA's recent proposals (Potter et al. 2021). It is believed that Titan, Ceres and perhaps even Pluto have subsurface liquid water oceans, which would considerably improve their habitability score, but we simply have no data, speculative or otherwise, that we can use to estimate their conditions in comparison to Earth's oceans. In future it is conceivable with increased data availability for other targets, the analysis of the ranking of each may change significantly. Although we cannot confirm that life exists on these targets, we can predict what type of life could survive under these conditions, as we have gleaned from this research.

In the case of Mars for example, the known radioresistant Earth bacterium called Deinococcus radiodurans can survive under the condition of radiation with a dose rate ~21 times higher than Mars (roughly equal to 5000 mSv). From a geochemical perspective, we have evidence of alteration minerals from chemical weathering that could be favourable for chemolithoautotrophic bacterium life also, like Acidophilic iron-sulphur bacterium and Acidithiobacillus ferrooxidans. In a model study, A. ferrooxidans was selected as a model organism to study its viability in the stimulating conditions of Mars. At the end of one week, it survived without any loss of viability under the stimulated conditions similar to that of shallow subsurface environments of Mars (-20°C temperature, 0.006 bar pressure and 0.18% of oxygen), and also helped in the reduction of minerals like goethite and hematite (Bauermeister et al. 2014).

Enceladus on the other hand, with its promising deep alkaline hydrothermal vents with an EHI score of 0.7778, is more suited to methanogenic life. When three thermophilic methanogenic strains were analysed, namely Methanothermococcus okinawensis, Methanothermobacter marburgensis, and Methanococcus villosus, the conditions of Enceladus were concluded to be favourable for their survival (Taubner et al. 2018). M. okinawensis showed more stability and reproducibility in additional extensive studies on deep subsurface conditions, such as Enceladus' ocean analogues (~50 bars, pH of ~5, >90°C).

Perhaps it is altogether not very surprising which bodies in the solar system have been deemed to have the greatest potential to support microbial life, with ocean worlds taking a dominant share of the spotlight. What is more surprising, however, is how this method has determined variation in habitability across different individual environments, to an extent which calls for more emphasis on particular fringe environment selection for future target exploration. When we consider the astrobiological potential of all our candidates, we compile a final list of habitability candidates in our solar system in the order of proposed likelihood to harbour life, and in addition, present our recommended order of priority for further astrobiological research and mission exploration. Finally, we have: Europa, Mars, Enceladus, Titan, Callisto, Ganymede and Pluto in order of habitability prospects. This selection is not totally comprehensive of course, but from our review of current research and literature, we have concluded that other moons, planets, asteroids, or other orbiting bodies in our solar system are less habitable - absolutely inhabitable in fact in nearly all cases, given the current bounds on microbial Earth life - than those which were included.

# 6. Conclusion

Many comprehensive studies have been conducted on microbial survivability for different planets and icy moons in the solar system, however there are not as many studies which focus on developing scoring systems and ranking methods for these astrobiological targets and their selected environments. Capitalising on this absence, in this research our team has reviewed the literature for some of the most studied astrobiological targets, filtered by data availability and bulk properties, and applied a new method and scoring system, called the Microbial Habitability Index (MHI), to assess and compare their relative propensity to support microbial life. The score value between 0 and 1 tells us how similar each environment is to its respective Earth analogue and also gives us information on the extent to which it is optimised to support microbial life. Combined with additional exogenic and non-numerical factors, our analysis yielded a final list of targets in order of microbial survival viability. From our results, much as expected, Europa, Mars and Enceladus take the top three positions, and we thus conclude them to be the best candidates to search for life in the solar system.

The main advantages of our method are apparent in its flexibility, both in terms of the environmental factors considered and the number of potential habitable environments. Even our somewhat limited inclusion of local variation in factors and environment types certainly is an improvement to a standard comparative approach of targets based on bulk properties. We must also emphasise that the considered habitability factors and environments are heavily influenced by the data availability. Looking towards future work, the technique can be expanded to incorporate more of both, as additional in-situ data is collected, and we learn more about other worlds in the solar system and their environments - especially sub-surface and water environments. The number of Earth fringe biospheres where life has been found is practically inexhaustible and is all the time increasing with discovery of new bacterial species and evidence of life in increasingly inhospitable niches, which too has bearing on how habitable an alien environment is deemed. External factors like long term geological changes, such as tectonic shifts, nutrient cycling, seasonal variations in many essential elements, as well as cosmic dangers and the endogenous changes have been mostly excluded from this analysis. We briefly mentioned in our background section the justification of this - and limiting our discussion to local environments at current time and over a sufficiently short time that the environmental conditions are stable and consistent, and that impact of these other longer-term changes are negligible. This does not mean however that these factors do not play a role in habitability over the entire history of a planet, and a more comprehensive study in the future would benefit from considering the full history of an astrobiological target, although this approach would present significant challenges.

Of course, much of our data is based on theoretical models - essentially all the data outside of that collected on Mars from on-site rovers, as well as from Enceladus, Europa and Titan attained by the Cassini/Huygens mission. Real on-site data is preferable and likely to improve accuracy, and we will have far more available to us in the near future, especially for distant moons and icy ocean worlds. Mars missions such as NASA's Mars 2020 Perseverance rover and CNSA's Zhurong rover, have the primary astrobiological goals to understand the evolution of Mars and search for traces of past life in their respective landing sites, namely Jezero crater and Utopia planitia. These next generation robots will be the first to give us more data essential for the kind of analysis done in our work, and future exploration will only aid us further. The upcoming ExoMars 2022 mission will explore the shallow subsurface (2 meters) environment for the first time. Aside from Mars, the rest of our solar system will be subject to similar explorations over the coming years. The Europa Clipper mission will study the surface, geology and ocean chemistry and also determine the course of formation of Europa's topographic features. The Jupiter Icy Moons Explorer (JUICE), will perform similar measurements for the Jovian moons Ganymede and Callisto. Even the routinely overlooked Titan will be explored by Dragonfly, a Rotorcraft lander mission which would study its surface composition and geology, investigate prebiotic chemistry and conduct a search for biosignature materials.

All these future proposed missions to the outer solar system targets would likely fill in many of the gaps that we currently have in our data and allow for a much more accurate analysis of environment based habitability. Many of our best environments by MHI score have crucial factors like temperature and UV-c radiation outside of the bounds for microbial habitability, and hence, maybe inhospitable to life. However, a positive point we would like to emphasise is that this method does have longevity due to its inherent scalability, in terms of all the relevant components, and can be amended by contemporary discoveries when the data becomes available, encompassing crucial microbial habitability factors, known potentially habitable environments, emerging limits of life, niche habitable Earth analogues determined by newly discovered extremophiles and even the choice of targets that can be assessed as habitable candidates; thereby broadening our horizons to the realm of exoplanets and beyond.

# Acknowledgments

This work was supported by the New York University Abu Dhabi (NYUAD) Institute Research Grant G1502 and the ASPIRE Award for Research Excellence (AARE) Grant S1560 by the Advanced Technology Research Council (ATRC). The authors would like to thank the Blue Marble Space Institute of Science (BMSIS) for facilitating this research and bringing together the collaborators through its Young Scientist Programme (YSP).